%% file: main.tex
\documentclass{article}
\usepackage{spconf,amsmath,graphicx}
\usepackage{caption}
\usepackage{algorithm}
\usepackage{algorithmic}
\usepackage{tabularx}
\usepackage{booktabs}
\usepackage{color,soul}
\usepackage{siunitx}
\usepackage{makecell}

\title{Minimum-Cost Sensor Channel Selection for Wearable Computing}
\name{Ramesh Kumar Sah$^{\dagger}$ \qquad Hassan Ghasemzadeh$^{\star}$ \thanks{This work was supported in part by the National Science Foundation under grant IIS-1954372. Any opinions, findings, conclusions, or recommendations expressed in this material are those of the authors and do not necessarily reflect the views of the funding organizations.}}
\address{$^{\dagger}$ Washington State University, $^{\star}$Arizona State University}
      
\begin{document}

\maketitle

\begin{abstract}
Sensor systems are constrained by design and finding top sensor channel(s) for a given computational task is an important but hard problem. We define an optimization framework and mathematically formulate the minimum-cost channel selection problem. We then propose two novel algorithms of varying scope and complexity to solve the optimization problem. Branch and bound channel selection finds a globally optimal channel subset and the greedy channel selection finds the best intermediate subset based on the value of a score function. Proposed channel selection algorithms are conditioned with performance as well as the cost of the channel subset. We evaluate both algorithms on two publicly available time series datasets of human activity recognition and mental task detection. Branch and bound channel selection achieved a cost saving of up to $94.8\%$ and the greedy search reduced the cost by $89.6\%$ while maintaining performance thresholds.



\end{abstract}


\input{sources/introduction}
\input{sources/problem_definition}

\input{sources/results}

\input{sources/conclusion}

\vfill\pagebreak
\bibliographystyle{IEEEbib}
\bibliography{refb}

\end{document}

%% file: sources/introduction.tex
\vspace{-2mm}
\section{Introduction}
\label{sec:introduction}
\vspace{-2mm}
Sensor systems usually operate in a restricted environment with limited computational and energy resources. Machine learning algorithms are increasingly used to make decisions in many applications of sensor systems \cite{ha2020machine}. Usage of machine learning and sensor systems is particularly prevalent and widespread in digital health applications \cite{sabry2022machine, triantafyllidis2019applications}. Often times many different sensing modalities are suitable to achieve a particular goal. For example, consider stress detection for which bio-markers such as heart rate variability \cite{dalmeida2021hrv}, skin conductance response \cite{sah2022stressalyzer}, and core body temperature \cite{herborn2015skin} are suitable. Another example could be human activity recognition, in which sensor systems can be placed at different body locations. In such cases, it becomes important to determine the optimal set of sensor channels to meet the requirements of the task while adhering to the design and operating limitations of the sensor system. 

Sensor channel selection is defined as the identification and removal of channels that provide a negative or negligible contribution to a goal task $T$. The problem is to select $k$ sensor channels out of $n$ total channels while optimizing the performance and total cost. Given this, there are $C(n, k) =  \frac{n!}{k! \times (n-k)!}$ channel subsets. For small search space, exhaustive search can be used to identify and remove redundant sensor channels. However, the search space grows exponentially with the size of channels set $(n)$ and exhaustive search is not feasible for large search spaces. A channel selection algorithm combines a search technique to find new channel subsets for evaluation and an evaluation method to evaluate the selected subset. A commonly used evaluation process involves training a machine learning algorithm for the considered task on the selected subset. The performance of the trained model is used as the proxy score of the selected channel subset. This approach of evaluation is similar to wrapper based feature selection extensively studied in machine learning literature. Most prior work on channel selection follows the wrapper based evaluation paradigm with some heuristic to either limit the search space of channel subsets \cite{espinilla2017optimizing, aziz2016identifying, ertuǧrul2017determining} or modifies the learning process to encourage the model to learn features from least number of input channels during the training process \cite{leite2021optimal, zappi2008activity, cao2018optimizing, yang2020instance}. In general these methods only consider the performance criteria in their evaluation step and the cost of channel subset is not used in the decision making process. This leaves gap in the research regarding optimal channel subset which not only meets the performance criteria but also has minimum total cost. 

In this work, we present two novel backward search algorithms to find a channel subset with minimum cost while ensuring a lower bound on the performance. The first algorithm, based on the branch and bound \cite{narendra1977branch} formulation, determines globally optimal channel subset and the second algorithm based on the greedy optimization selects the best intermediate subsets based on the value of a score function. Branch and bound yields a set of channel subsets that meets the performance threshold and the greedy search returns a single channel subset. Our proposed channel selection algorithms consider both the performance and the cost of the subset in the evaluation step. Furthermore, the branch and bound channel selection returns globally optimal channel subset. We validate the proposed channel selection algorithms with two publicly available time-series datasets. 

%% file: sources/problem_definition.tex
\vspace{-2mm}
\section{Minimum Cost Channel Selection}
\label{sec:problem_definition}
\vspace{-2mm}
Minimum cost channel selection (MCCS) is defined as identifying and selecting a subset of channels out of $n$ channels while ensuring a lower bound ($\lambda$) for the performance function $f$ and minimum total cost $W$. 
Furthermore, depending on the machine learning task $T$ the performance function $f$ can either be maximized (accuracy, f1-score) or minimized (mean squared error). 

\vspace{-2mm}
\subsection{Problem Definition}
Given $n$ sensor channels $C_n = \{c_1, c_2, \dots c_n\}$ and cost $W_n = \{w_1, w_2, \dots, w_n\}$ for selecting each sensor channel. The MCCS problem is to minimize the total cost of the selected subset $C_s$
\begin{equation}
    \text{minimize} \sum_{i=1}^{s}{w_i}    
\end{equation}
subject to 
\begin{equation}
    f(C_s) \geq \lambda
\end{equation}

Here, $f$ is a performance function, $\lambda$ is the lower bound on the performance, and $w_i$ is the normalized cost of selecting channel $c_i$. Normalized cost is obtained for all channels given $W_n$ such that $\sum_{i=1}^{n}{w_i} = 1$.




\vspace{-2mm}
\subsection{Branch and Bound Channel Selection}
Let $(c_1, \dots, c_{\Bar{s}})$ be the $\Bar{s} = n - s$ channels to be discarded to obtain the channel subset $(C_s)$ of size $s$. Each channel $c_i$ can take on value in $\{1, 2, \dots, n\}$. Here, the order of $c_i$'s is not important, and we only consider sequences of $c_i$'s such that $c_1 < c_2 < \dots < c_{\Bar{s}}$. The performance function $(f)$ is a function of the selected channel subset $(C_s)$ obtained by discarding $c_1, \dots, c_{\Bar{s}}$ channels from the $n$ channel set. Now, the channel subset selection problem is to find the subset $c_1^*, \dots, c_{\Bar{s}}^*$ to discard such that

\begin{equation} \label{BB_Equation}
    \begin{split}
        & f(C_{\Bar{s}}^*) = \text{max } f(C_s^*) \\ 
        \text{and } & W(C_{\Bar{s}}^*) = \text{min } W(C_s^*)
    \end{split}
\end{equation}

\noindent
$W$ is a cost function defined as the sum of the normalized cost of all channels in the selected subset $C_s$. Let us assume the performance function $f$ satisfies monotonicity defined by 
\begin{equation} \label{Performance_Monotonicity_Equation}
    f_n(c_1, c_2, \dots, c_n) \geq f_{n-1}(c_1, c_2, \dots, c_{n-1}) \geq \dots \geq f_1(c_1)  
\end{equation}

\noindent
The monotonicity principle means that a subset of channels should not be better than any larger set containing the subset. We acknowledge that not all types of neural networks satisfy the monotonicity principle, but recent works have shown ways to create deep neural networks with monotonic properties \cite{runje2022constrained}. The cost function already satisfies the principle of monotonicity i.e., $W_n(c_1, c_2, \dots, c_n) \geq W_{n-1}(c_1, c_2, \dots, c_{n-1}) \geq \dots \geq W_1(c_1)$. Then, given the lower bound $(\lambda)$ on the value of the performance, we can write 
\begin{equation}
    \lambda \leq f(C_s^*)
\end{equation}

\noindent
And, if $f(C_k) (k > s)$ is less than $\lambda$, then from equation \ref{Performance_Monotonicity_Equation},

\begin{equation} \label{BB_Criterion_Equation}
    \begin{split}
        & f(C_s) \leq \lambda \\
        & \forall \quad \{C_{k+1}, \dots, C_s\}
    \end{split}
\end{equation}

\noindent Equation \ref{BB_Criterion_Equation} means that whenever the performance function evaluated for any subset is less than $\lambda$, all subsets that are successors of that subset also have performance value less than $\lambda$, and therefore cannot be the optimum solution. This form the basis for the branch and bound channel selection algorithm. Branch and bound method successively generates portions of the solution tree and computes the performance value. Whenever a sub-optimal partial subset satisfies condition \ref{BB_Criterion_Equation}, the sub-tree under that subset is implicitly rejected, and enumeration begins on the subsets which have not yet been evaluated \cite{narendra1977branch}. Algorithm \ref{alg:branch_and_bound_algorithm} describes the proposed branch and bound channel selection.

\begin{algorithm}[!tbh]
\small
    \caption{Branch and bound channel subset selection}
    \label{alg:branch_and_bound_algorithm}
    \textbf{Input}: List of channels $C_n = \{c_1, c_2, \dots, c_n\}$, Cost of each channel $W_n = \{w_1, w_2, \dots, w_n\}$, and Number of channels $n$\\
    \textbf{Parameter}: Objective function $f$ and Performance threshold $\lambda$\\
    \textbf{Output}: Globally optimal channel subset $C^*$, cost of the selected channel subset $W^*$, and list of optimal subsets $C_o$
    \begin{algorithmic}[1] 
        \STATE Set $C^* = C_n$ and $W^* = \sum{}{}{w_i}$
        \STATE Create stack $S$ and hash table $H$
        \STATE Set current subset node $K_{current}$ = $C_n$
        \STATE $C_o$ = []
        \STATE POPPED $= 1$
        \IF {$f(K_{current}) < \lambda$}
        \STATE \textbf{return} $C^*$, $W^*$, $C_o$
        \ENDIF
        \STATE Push $K_{current}$ into $S$
        \STATE Map $K{current}$ into $H$
        \WHILE{$S$ is not empty}
        \STATE $K_{previous} = K_{current}$
        \STATE Create children subset nodes of $K_{current}$ and store them in the ascending order of the cost in $L$
        \FOR{subset $n$ in $L$}
        \STATE $K_{current} = n$
        \STATE Check the performance of $K_{current}$ and update $S$, $H$
        \STATE Update $C^*$, $W^*$ and $C_0$ if needed
        
        \ENDFOR
        \IF{$K_{current} == K_{previous}$}
        \STATE Pop S and assign to $K_{current}$
        \STATE POPPED $ = 1$
        \ELSE
        \STATE POPPED $ = 0$
        \ENDIF
        \ENDWHILE
        \STATE \textbf{return} $C^*$, $W^*$, $C_o$
    \end{algorithmic}
\end{algorithm}

\vspace{-6mm}
\subsection{Greedy Channel Selection}
The branch and bound algorithm assume monotonicity in performance which may not be always true. Furthermore, in the worst case branch and bound search must evaluate all possible channel subsets and consequently will have an exponential runtime \cite{zhang1996branch}. In light of these limitations, we also propose a greedy Algorithm \ref{alg:greedy_algorithm} for sub-optimal channel subset selection. 

Let $C$ be a root channel subset node and $C-c$ be its children subset node. The subset $C-c$ is created by discarding the channel $c$ from the parent subset $C$. We define a \textit{score} function 
\begin{equation} \label{Score_Equation_Original}
    \text{score} = \alpha \times f(C-c) + (1 - \alpha) \times W(C-c)
\end{equation}

\noindent
where $f(C-c)$ is the value of the performance function on the channel subset $C-c$ and $W(C-c)$ is the sum of the normalized cost of channels in the subset $C-c$. $\alpha$ is a balancing term used to control the influence of performance and cost on the score value. Given the score function, the greedy algorithm selects a channel subset that meets the performance threshold i.e., $f(C-c) \geq \lambda$ and has the minimum value of score at each intermediate stage. Also, since the goal is to minimize the value of the score function, for classification problems we modify the score function as 

\begin{equation} \label{Score_Equation_Modified}
    \text{score} = \alpha \times (1 - f(C-c)) + (1 - \alpha) \times W(C-c)
\end{equation}
The algorithm greedily selects a subset with the least score value and hence is able to achieve a runtime of $O(n^2)$.

\begin{algorithm}[!tbh]
\small
    \caption{Greedy channel subset selection}
    \label{alg:greedy_algorithm}
    \textbf{Input}: List of channels $C_n = \{c_1, c_2, \dots, c_n\}$, Normalized cost of each channel $W_n = \{w_1, w_2, \dots, w_n\}$\\
    \textbf{Parameter}: Objective function $f$, Performance threshold $\lambda$, and $\alpha$\\
    \textbf{Output}: Locally optimal channel subset $C$ and cost $W$
    \begin{algorithmic}[1] 
        \STATE Set $C = C_n$ and $W = \sum{}{}{w_i}$
        \STATE Set current subset node $C_k$ = $C$
        \IF {$f(C_k) < \lambda$}
        \STATE \textbf{return} $C$, $W$
        \ENDIF
        \STATE Set best score $S_b = inf$
        \WHILE{true}
        \STATE $L = $ children subset nodes of $C_k$
        \IF{L is Empty}
        \STATE \textbf{return} $C$, $W$
        \ENDIF
        \FOR{subset S in L}
        \STATE Check performance of S
        \STATE Update $C$ and $W$ if needed
        \ENDFOR
        \ENDWHILE
        \STATE \textbf{return} $C$, $W$
    \end{algorithmic}
\end{algorithm}

\vspace{-4mm}
\subsection{Cost Model}
One important aspect of channel subset selection is the cost associated with each sensor channel. We propose to use a cost model to obtain the cost for a channel. Cost model defines the cost of a channel based on the some input parameters such as: 1) computation and memory requirement which are directly related to sampling frequency, 2) power requirement, 3) sensing requirement, 3) usability and interpretability cost, 4) manufacturing cost, and 5) other cost. In our analysis, we generate the cost for each channel using a simple heuristic based on the sampling frequency of the sensor channel. Sensor channel with higher sampling frequency are assigned a larger cost and vice-versa. In practice, the cost of the sensor channel can be determined as needed and used with our proposed algorithms to determine the optimal channel subset.




%% file: sources/results.tex
\vspace{-3mm}
\section{Analysis and Results}
\label{sec:results}
\vspace{-2mm}

\subsection{Datasets}
\textit{EEG Mental Task} \cite{zyma2019electroencephalograms} dataset contains electroencephalogy (EEG) signals recorded for binary mental arithmetic task detection using Neurocom EEG $23$-channel system at sampling frequency of $500$ Hz. Electrodes were attached to the cranium at certain regions of symmetrical anterior frontal, frontal, central, parietal, occipital, and temporal sites. A high-pass filter with a $30$ Hz cut-off frequency and a power line notch filter ($50$ Hz) were used to eliminate noise and artifacts from all EEG channels. All recordings are artifact-free EEG segments of $60$ seconds duration. We further subdivided the EEG segments into input windows of size $10$ seconds with $5$ seconds overlap between consecutive windows. 

\textit{PAMAP2} \cite{reiss2012introducing} is a human activity recognition dataset recorded from $9$ participants wearing $3$ sensor units and performing $18$ activities. Each sensor unit contained $3-$axis accelerometer, gyroscope, and magnetometer all sampled at $100$ Hz. Sensor units were placed at chest, wrist, and ankle of the participants. Altogether we have $27$ sensor channels with $9$ channels from each body location. Signal from each sensor is subdivided into windows of size $30$ seconds with $15$ seconds overlap between consecutive windows. The activity recognition task is defined as a $7$ class classification problem. Please consult the original publications for more details about the sensor channels in both datasets \cite{zyma2019electroencephalograms, reiss2012introducing}.



\begin{figure}[tbh]
    \centering
    \includegraphics[width=\linewidth]{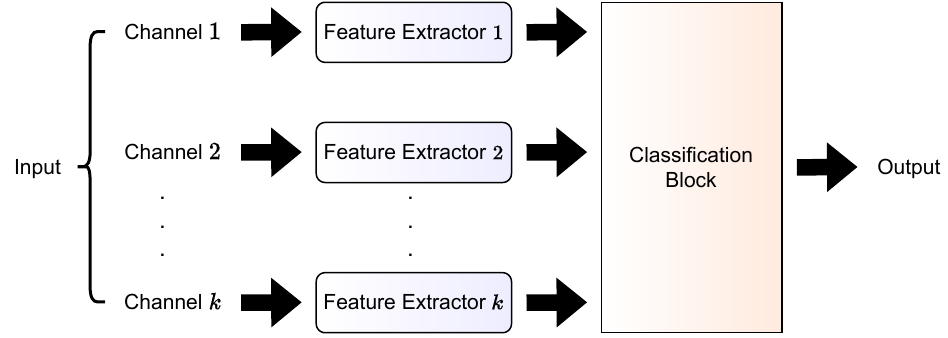}
    \caption{Modular architecture of the CNN model with number of feature extractor equal to the number of input channels.}
    \label{fig:cnn_block}
\end{figure}

\vspace{-8mm}
\subsection{Model Architecture}
We have used $1D$ Convolutional Neural Network (CNN) architecture to evaluate each channel subset during the search process. CNNs are known to work well for time-series classification problems \cite{zeng2014convolutional, sah2022stressalyzer} and can be trained with raw sensor values without feature computation and selection. CNN learns the feature and mapping between input and output during the training process. During evaluation the number of input channels to the model depend on the size of the evaluation channel subset. Hence, we have used a modular architecture of CNN capable of accommodating dynamic change in input channels as shown in Fig \ref{fig:cnn_block}. Each input channel in the considered subset is assigned a separate feature extraction block and outputs from all feature extractors is aggregated in the classification block to learn the mapping from input to output. The feature extraction block consists of two one-dimensional convolutional layers and the classification block has two fully-connected layers. ReLU activation is used in all intermediate layers and Softmax activation in the output layer consisting same number of neurons as the number of output classes. In all cases, the model is trained for $100$ epochs using cross entropy loss and Adam optimizer with learning rate set to $0.001$. 

\vspace{-2mm}
\subsection{Channel Subset Selection}
\vspace{-1mm}
In our experiments, we initially set $\alpha = 0.5$ for greedy channel search and measured the performance in terms of accuracy of the trained model. Table \ref{tab:res_eeg} shows the optimal channel subset for EEG mental task dataset determined using branch and bound and greedy channel selections. Since the sampling frequency of all channels in the EEG dataset is equal, the normalized cost of each channel is also equal and set to $w_i = 0.052$. The accuracy of the model trained on all available channels is considered baseline performance, and for the EEG dataset the baseline accuracy was $73.48\%$. Given the baseline performance, the performance threshold $\lambda$ was set to $0.7$ or $70\%$ accuracy. For EEG dataset, branch and bound channel selection was able to achieve a cost saving of $94.8\%$ and the greedy search was able to reduce the cost by $89.6\%$.



\vspace{-1mm}
\begin{table}[tbh]
    \caption{Optimal channel subset for EEG dataset determined using branch and bound (B\&B) and greedy methods.}
    \label{tab:res_eeg}
    \setlength\tabcolsep{5.3pt}
    \begin{tabular}{ccccccc}
        \toprule
        \multicolumn{1}{c}{Method} & \multicolumn{1}{c}{Selected} & \multicolumn{1}{c}{Accuracy} & \multicolumn{1}{c}{Cost} & \multicolumn{1}{c}{Score} & \multicolumn{1}{c}{Cost}\\
        {} & {Subset} & {$(\%)$} & {} & {} & {Savings}\\
        \midrule
        B\&B & FP1 & 70.31 & 0.052 & 0.17 & $94.8\%$\\
        Greedy & (C3, F3) & 72.33 & 0.104 & 0.19 & $89.6\%$\\
        \bottomrule
    \end{tabular}
\end{table}
\vspace{-1mm}

\noindent
All channels in the PAMAP2 dataset also has equal sampling frequency, and consequently the cost of each channel is equal and set to $w_i = 0.037$. The baseline performance accuracy was determined to be $59.02\%$ and the performance threshold $\lambda$ was set to $0.5$ or $50\%$ accuracy. The subset (CA2, AG3) was found to be globally optimal using branch and bound search with performance $51.22\%$, cost $0.074$, and score $0.282$. The subset (HA1, HA3, CA1, CA2, CA3, CG2, CG3, CM1, AA1, AA2, AA3, AG1, AM1) was determined to be best using greedy search. The cost of greedy subset is $0.4814$ with performance of $89.75\%$ and score of $0.2919$. A cost saving of $92.6\%$ was achieved with branch and bound search and a cost saving of $51.8\%$ with greedy search.

\vspace{-2mm}
\subsection{Effects of Alpha}
\begin{figure}[tbh]
    \centering
    \includegraphics[width=\linewidth]{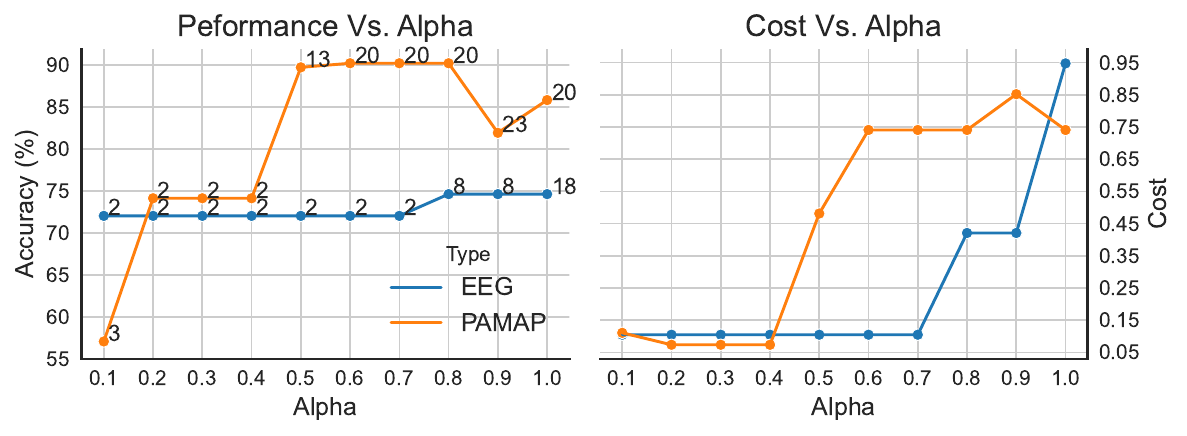}
    \caption{Accuracy and cost of the selected channel subset using greedy search for both datasets at different values of alpha. The values on the line denotes the number of selected channels. For example at $\alpha=0.6$, $2$ channels were selected for the EEG mental arithmetic task and $20$ channels were selected for PAMAP activity recognition task.}
    \label{fig:alpha_var}
\end{figure}

\vspace{-1mm}
We set $\alpha = 0.5$ in the preceding analysis placing equal importance on the cost and performance for greedy channel selection. However, in practice minimizing cost might be more important than maximizing performance and vice-versa. Fig \ref{fig:alpha_var} shows the performance and cost of the selected channel subset at different values of $\alpha$. Larger values of $\alpha$ puts greater emphasis on the performance and smaller values of $\alpha$ favours channel subset with lower costs. For both datasets, at larger values of $\alpha$ the accuracy of the selected subset is higher but the cost is also high. This is expected because greater number of input channels will provide more (most likely better) information to the model to learn the mappings between input and outputs, consequently increasing the performance. 

\vspace{-2mm}
\vspace{-2mm}

%% file: sources/conclusion.tex
\vspace{-5mm}
\section{Conclusion}
\label{sec:conclusion}
\vspace{-3mm}
In this work, we have proposed and validated two sensor channel selection algorithms to determine optimal subset of channels that meets the performance criteria with minimum cost. Proposed algorithms can be used in real-life applications to optimize the cost of a sensor system while also ensuring a performance guarantee. Branch and bound channel selection also allows for dynamic selection of channels during run-time since it returns a list of channel subsets satisfying the performance threshold. When some channels from the globally optimal subset becomes unavailable during run-time, channels from next best subsets can be used to keep the system operational. Our evaluation scheme is model agnostic and any other type of learning algorithm can be used instead of CNN. We also note that training separate models for each subset during evaluation might not be feasible for a large search space. In these cases, a single model can be trained on all available channels and during search channels that are absent from the evaluation subset can be masked out \cite{leite2021optimal}. Training a single model also allows for the dynamic selection of channels during run-time based on the availability.